\documentclass[12pt]{iopart}
\usepackage{epsfig}
\usepackage{citesort}
\usepackage{multicol, color}

\newcommand{\gtrsim}{\,\rlap{\lower3.7pt\hbox{$\mathchar\sim$}}
\raise1pt\hbox{$>$}\,}
\newcommand{\lesssim}{\,\rlap{\lower3.7pt\hbox{$\mathchar\sim$}}
\raise1pt\hbox{$<$}\,}

\newcommand{\be}{\begin{equation}}
\newcommand{\ee}{\end{equation}}  
\newcommand{\bea}{\begin{eqnarray}}
\newcommand{\eea}{\end{eqnarray}}  
\newcommand{\gag}{g_{a\gamma}}

\begin{document}
\hfill   MPP-2009-145

\title{Stochastic conversions of TeV photons into axion-like  particles in 
extragalactic magnetic fields $ $ $ $ $ $}

\author{Alessandro Mirizzi $^{}$
\footnote{$^{}$ Current address: II. Institut f\"ur theoretische Physik, Universit\"at Hamburg, 
Luruper Chausse 149, 22761 Hamburg, Germany.}}
\address{
Max-Planck-Institut f\"ur Physik (Werner Heisenberg
Institut) \\ 
F\"ohringer Ring 6, \\
80805 M\"unchen, Germany}
\author{Daniele Montanino}
\address{Dipartimento di Fisica, Universit\`a del Salento\
and Sezione INFN di Lecce\\
Via Arnesano, \\
I--73100 Lecce, Italy
}

\begin{abstract} 
Very-high energy photons emitted by distant cosmic sources are absorbed
on the extragalactic background light (EBL) during their propagation.
This effect can be characterized in terms of a photon transfer function at Earth. 
The presence of  extragalactic magnetic fields could also induce
conversions between very high-energy photons and hypothetical axion-like particles (ALPs).   
The turbulent structure of the
 extragalactic magnetic fields  would produce a stochastic behaviour in these conversions,
 leading to a statistical distribution of the photon transfer functions for the different
 realizations 
 of the random magnetic fields.
To characterize this effect, we derive new equations to
calculate  the mean and the variance of this  distribution.
 We find  that, in presence of ALP
conversions, the photon transfer functions on different lines of sight  could have relevant deviations with respect to the mean value, producing
 both an enhancement or a   suppression in the observable photon flux
 with respect to the
expectations with only absorption. As a consequence, the most striking signature of the mixing  with
ALPs would be a reconstructed EBL density
from TeV photon observations which appears to vary over different directions
of the sky: consistent
with standard expectations in some regions, but inconsistent in others. 
\\
\noindent {\em Keywords}: axions, very high-energy gamma-rays. 

\end{abstract}

\maketitle

\section{Introduction}

Axion-like particles (ALPs) with a two-photon vertex are predicted in many
extensions of the Standard Model~\cite{Svrcek:2006yi,Arvanitaki:2009fg,Masso:2006id}. 
The $a\gamma\gamma$ coupling allows for ALP-photon conversions in electric
or magnetic field. 
This effect  is exploited by the ADMX experiment to search for   axion dark 
matter~\cite{Duffy:2006aa},
 by CAST to search for 
solar axions~\cite{Zioutas:2004hi,Andriamonje:2007ew,Arik:2008mq}, 
and by regeneration
laser experiments~\cite{Robilliard:2007bq,Chou:2007zzc,%
Fouche:2008jk,Afanasev:2008jt,ringwald}. 

ALPs  also play an intriguing role in astrophysics. 
Indeed,   photons emitted by distant 
sources and propagating  through  cosmic magnetic fields
can oscillate into ALPs.  The consequences of 
this effect have been studied in different 
situations~\cite{Csaki:2001yk,Mirizzi:2005ng,Mirizzi:2006zy,Mirizzi:2009nq,Csaki:2003ef,Mirizzi:2007hr,Dupays:2005xs,%
Hooper:2007bq,De Angelis:2007yu,Fairbairn:2009zi,Burrage:2009mj}.
 In particular, in the last recent years
photon-ALP conversions have been proposed as a mechanism to 
avoid the opacity
of the extragalactic sky to high-energy radiation due to
pair production on the Extragalactic Background Light
(EBL). 
 At this regard, recent observations of cosmologically distant gamma-ray 
sources by ground-based gamma-ray telescopes have
 revealed a surprising degree of transparency of the universe
 to very high-energy (VHE) photons ($E\gtrsim 100$~GeV)~\cite{Aharonian:2005gh,Mazin:2007pn}.
Surprisingly,  data seem to require a lower density of the EBL than expected
and/or considerably harder injection spectra than initially
thought~\cite{Stecker:2007jq,Stecker:2007zj}. 
Oscillations between very high-energy  photons and ALPs could represent an 
intriguing possibility to explain this puzzle 
through a sort of
``cosmic light-shining through wall'' effect.
Infact, if VHE  photons are converted into ALPS and then regenerated,
 they should not suffer absorption effects while they propagate as ALPs.
In this sense, two complementary mechanisms have been proposed:
a)  VHE photon-ALP conversions in the magnetic fields
 around gamma-ray sources~\cite{Hooper:2007bq,Hochmuth:2007hk} and then  further
back-conversions in the magnetic 
field of the Milky Way~\cite{Simet:2007sa} (see also~\cite{Bassan:2009gy}); 
b)  
oscillations of VHE photons into ALPs in
the random extragalactic magnetic 
fields~\cite{De Angelis:2007dy,DeAngelis:2008sk}. 
In principle, both the mechanisms can be combined together, as 
shown in~\cite{SanchezConde:2009wu}.
Currently, the inference of EBL from   VHE photons emitted
by sources at high redshift~\cite{alberto} is still object of debate, and
it is not clear how robust are the conclusions 
on the absorption effects obtained  from the recent gamma data~\cite{Costamante:2009gz}.
In this sense, it is also possible that the observed 
transparency of the universe to VHE photons could be explained
without the need of introducing nonstandard mechanisms (see, e.g.,~\cite{Aharonian:2008su,Essey:2009zg}). 
Nevertheless, the seminal works mentioned before have pointed out the nice 
connection between VHE gamma-astronomy and ALP searches.
Therefore, it seems  worthwhile to further explore  the 
consequences of this exciting possibility.    

In this context, the treatment of the oscillations of VHE photons into
ALPs in the extragalactic medium in presence  of the  absorption
on the EBL presents a certain degree of complexity. Extragalactic
magnetic fields are supposed  
 to have a turbulent structure which can dramatically affect the development
of photon-ALP conversions.
 Brute force numerical simulations  get
the solution of the mixing equations  along
a given photon line of sight by iterating the 
equations in each domain in which the magnetic field is
 assumed constant~\cite{Csaki:2003ef,De Angelis:2007dy}. 
Since one cannot know the given
configuration of magnetic domains crossed by VHE photons during their
propagation, 
in the previous literature VHE photon-ALP conversions
were usually characterized in terms of the mean conversion probability, obtained
 averaging the resulting conversion probabilities 
over an ensemble of magnetic field configurations along the photon line of sight.
Such a procedure  slows down  the solution of our problem, since
to obtain stable results
typically the average has to be performed over more than $10^{3}$ realizations
of the magnetic fields~\cite{De Angelis:2007dy}.
Moreover, the use of the mean probability as representative value does not appear
 a priori completely justified, since 
the variance of the probability distribution
could produce relevant deviations from the mean value for  conversions occurring  in different
realizations of the random magnetic fields.

In order to overcome these previous limitations, in this paper we perform
a new study of the mixing equations of VHE photons in the turbulent magnetic fields and we  
provide a user-friendly calculation  of the mean and of the
variance for the distribution of the photon transfer functions. 
The plan of our work is as follows.
In Section~2 we  discuss about the  absorption of VHE photons 
on the extragalactic background light.
In Section~3 we  characterize  VHE photon-ALP mixing in presence
of absorption and we  present our calculation
of the mean photon transfer function averaged
over the ensemble of all the possible realizations of random  magnetic field 
configurations.
We also show how to calculate the variance of the statistical distribution of the transfer functions.
In Section~4 we  present our results for
the photon transfer function
of  VHE gamma-rays emitted from distant sources
with and without the mixing with  ALPs.
Contrarily to  previous predictions,  the presence of a  broad  variance 
in the statistical distribution of the photon transfer functions   
could produce both an enhancement or a suppression 
of the observed VHE photon flux with respect to the case with
only absorption. The resulting photon flux would depend on the particular
random magnetic field configuration along the photon  line of sight.
As a consequence, photon-ALP mixing can not provide
an universal mechanism to obtain the transparency of the universe
to VHE radiation, but instead they would produce a strong direction-dependent
 behaviour in the flux of VHE photons from distant sources.
 In Section~5 we  discuss about possible developments
of our study and we conclude. There follow two Appendices, in which we
present some details for the derivation of the mean photon transfer function 
(Appendix~A) and for the variance of the distribution
 (Appendix~B).

%........................................................
\section{Absorption of very high-energy photons on extragalactic background light}
%......................................................

The flux of very high-energy (VHE) gamma rays ($E\gtrsim 100$~GeV) from distant sources 
is attenuated in an energy dependent way by the interaction with 
background photons in the universe. 
The main source of absorption for VHE photons 
 is due to the pair production process $\gamma^{\rm VHE}\gamma^{\rm bkg}\to e^+e^-$.
 In the energy range 100~GeV $\lesssim E \lesssim$ 10 TeV, 
 the absorption is dominated
 by the interaction with optical/infrared photons 
of the so called
 Extragalactic Background Light (EBL), 
sometimes also referred  as Metagalactic Radiation Field (MRF).
 The absorption rate  for such a process
 in function of  the incident photon  energy $E$
is given by~\cite{Gould:1967zzb}
\begin{equation}
\Gamma_\gamma(E)=\int_{m_e^2/E}^{\infty} d\epsilon\, \frac{dn^{\rm bkg}_\gamma}{d\epsilon}
\int_{-1}^{1-\frac{2m_e^2}{E\epsilon}} d\xi\frac{1-\xi}{2}\sigma_{\gamma\gamma}(\beta) \,\ ,
\, \label{eq:gamma}\end{equation}
where the limits of integration in both integrals are determined by the kinematical threshold
of the process and
\begin{equation}
\nonumber
\sigma_{\gamma\gamma}(\beta) = \sigma_0
(1-\beta^2)\left[2\beta(\beta^2-2)+(3-\beta^4)\log\frac{1+\beta}{1-\beta}\right]
\, ,
\end{equation}
with $\sigma_0=1.25\times 10^{-25}$ cm$^2$, is the cross section for the pair production process~\cite{Heitler:1960},
in function of the electron velocity in the center of mass of the interaction
$\beta=[1-2m_e^2/E\epsilon(1-\xi)]^{1/2}$, 
being $\epsilon$ the background photon energy, and 
$\xi$ the cosine of the angle between the incident 
and the background photon.
For practical purposes, it can be useful to notice that
the inner integral in $d\xi$  in Eq.~(\ref{eq:gamma}) can be evaluated by performing the
change of variable $\xi\equiv\xi(\beta)$ and has actually an analytic closed form:
\begin{eqnarray}
&\phantom{=}& 4\sigma_0 (1-\beta_m^2)^2\cdot\left[
{\rm Li}_2\left(\frac{1-\beta_m}{2}\right)-
{\rm Li}_2\left(\frac{1+\beta_m}{2}\right)\right.
-\frac{\beta_m(1+\beta_m^2)}{1-\beta_m^2}\nonumber\\
&\phantom{=}&\left.+\frac{1}{2}\left(\frac{1+\beta_m^4}{1-\beta_m^2}-\log\frac{1-\beta_m^2}{4}\right)
\log\frac{1+\beta_m}{1-\beta_m}
\right]\, ,
\end{eqnarray}
where $\beta_m=(1-m_e^2/E\epsilon)^{1/2}$ is the maximum electron velocity 
in the center of mass of the interaction and the function ${\rm Li}_2$ 
is the polylogarithm of order two. 

%%%%%%%%%%%%%%%%%%%%%%%%%%% FIGURE 1 %%%%%%%%%%%%%%%%%%%%%%%%%%%%%%%%%
\begin{figure}[!t]
\centering
\epsfig{figure=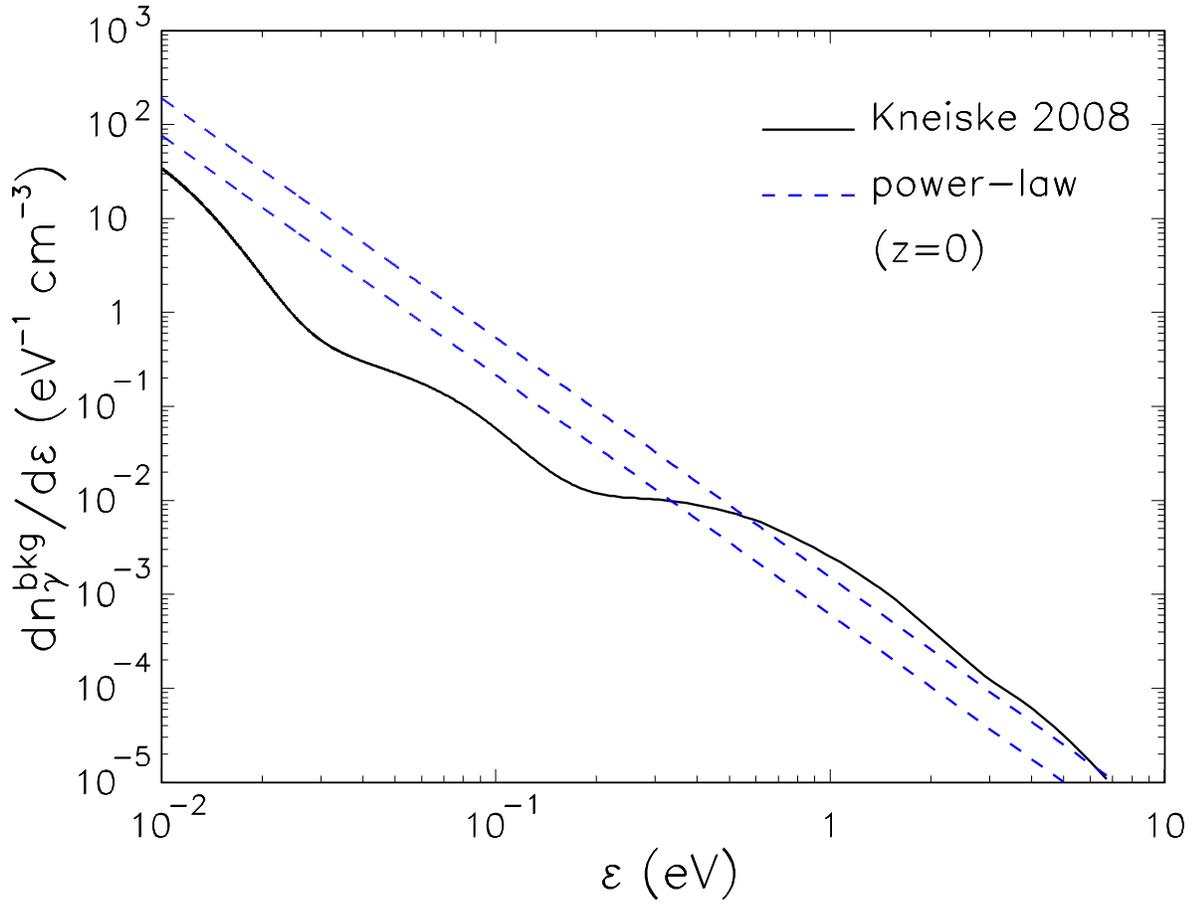, width =1.0\columnwidth, angle=0}
 \caption{Spectrum of  the EBL relevant for the absorption
of VHE photons. The dashed power-law lines correspond
to the simplified model of Eq.~(\ref{eq:bgktoday}) 
for the limiting cases of $k=0.61$ (lower line)
and $k=1.52$ (upper line) respectively. 
The solid line corresponds to the 2008 Minimal Kneiske Model
at redshift $z=0$. (see the text for details)
\label{ebl}} 
\end{figure}
%%%%%%%%%%%%%%%%%%%%%%%%%%%%%%%%%%%%%%%%%%%%%%%%%%%%%%%%%%%%%%%%%%%%% 

For simple estimations, the background photon spectrum 
in Eq.~(\ref{eq:gamma})
 can be approximated, at redshift $z=0$, with  a power--law~\cite{Salamon:1994un}
\begin{equation}
\frac{dn^{\rm bkg}_\gamma}{d\epsilon}=10^{-3}k
\left(\frac{\epsilon}{{\rm eV}}\right)^{-2.55}\,{\rm eV^{-1}}\,{\rm cm}^{-3}\,,
\label{eq:bgktoday}\end{equation}
with $0.61\leq k\leq 1.52$, depending on the model used 
(see Fig.~\ref{ebl}). With this approximation one has  
\begin{equation}
\frac{\Gamma_\gamma(E)}{{\rm Mpc}^{-1}}\simeq 1.1\times 10^{-3}\,k\left(\frac{E}
{\rm TeV}\right)^{1.55}\,.\label{eq:lenght}
\end{equation}
In the literature are present  
different realistic models for the photon background. 
In particular, in the following 
we will refer to the  2008 Minimal Kneiske Model~\cite{Kneiske:2008mp}, which
 provides 
 a strict lower-limit flux for the extragalactic background light from 
ultraviolet to the far-infrared photon energies.
  The model parameters are 
chosen to fit the lower limit data from galaxy number count 
observations,   assuming that the shape and the normalization
of the background radiation does not change except for the red-shifting
between the source and the observer.
This model has  the advantage that  it is not inferred by an inversion of 
the VHE photon observed spectra, as e.g. in~\cite{Mazin:2007pn}.
We note that  this latter procedure  would  be not 
reliable in the presence of   VHE photon-ALP conversions,
since in this case the effects of mixing and absorption on EBL would be entangled,
preventing  the possibility to extract
information on the EBL from VHE photon measured spectra. 
Our reference model gives us the maximal possible transparency compatible with the
standard expectations, so that an evidence of a greater transparency would
have to be attributed to  nonstandard effects in the photon propagation.

The spectral energy distribution at a given redshift $z$
 can   be inferred by the tabulated power spectrum~\cite{Kneiske:2008mp}
 $P(\lambda,z)=\lambda I_\lambda(\lambda,z)$ (where $I(\lambda,z)$ 
is the flux at redshift $z$ of energy for unit of solid angle
 between $\lambda$ and $\lambda+d\lambda$ where $\lambda$ is the comoving wavelength) 
by 
\begin{equation}
\frac{dn^{\rm bkg}_\gamma}{d\lambda}=2P(\lambda(1+z))(1+z)^3\, .
\end{equation}
The corresponding EBL energy spectrum is represented in Fig.~1. 
 From this Figure we see that the realistic Kneiske model
 has an approximate power-law trend 
at photon background energies $\epsilon \gtrsim$~few~$\times 10^{-1}$~eV, which are
relevant for the absorption of TeV photons. This will be useful to explain the high-energy behaviour of 
 the  transfer function. 

The absorption function $\Gamma_\gamma$ as function of the photon energy on Earth 
$E$ and of the redshift $z$
 is  given by Eq.~(\ref{eq:gamma}) simply changing
 $E \to E(1+z)$ in the integrand and in the limit of integration.

Finally, for a given source at distance $L$, the photon spectrum observed on Earth
(apart from the geometrical dilution) is  given by%
\begin{equation}
 I_{\rm obs}(E)
=T_\gamma(E,L)\cdot I_{\rm source}(E_0) = \exp\left(-\tau_\gamma\right)
I_{\rm source}(E_0) \, ,
\label{eq:noconversion}\end{equation} 
where 
the initial source spectrum 
$I_{\rm source}(E_0)$,  with initial photon energy $E_0=E(1+z)$, is modified due to the
effect of the VHE photon absorption, through the transfer function  $T_\gamma$, expressed in terms of
 the {\em optical depth}
%................................
\begin{equation}
\tau_\gamma=\int_0^L dx \,\Gamma_\gamma(E,x)= \frac{c}{H_0}\int_{0}^{z_0}\frac{dz}{(1+z)
\sqrt{\Omega_\Lambda + \Omega_m(1+z)^3}}\Gamma_{\gamma}(E,z) \, ,
\end{equation}
%%%%%%%%%%%%%%%%%%%%%%%%%%%%%%%%%%%%%%%%%%%%%%%%%%%%%%%%%%%%%%%%%%%%%%%%%%%%%
dependent on the evolution of the Universe through  the Hubble constant $H_0=73$~km~Mpc$^{-1}$~s$^{-1}$, 
the matter density $\Omega_m=0.24$~\cite{Amsler:2008zzb}, and the 
 dark energy density   $\Omega_\Lambda =1-\Omega_m$ (assuming a flat cosmology).
We observe that in absence of photon-ALP conversions the transfer function $T_\gamma$
 drops exponentially at high energies. We will see that the effect of
 VHE photon-ALP conversion is to soften this behaviour.

%....................................
\section{Very high-energy photons mixing with axion-like particles}
%................................

\subsection{Equations of motion}

The effect of the absorption of VHE photons on the EBL,
described in the previous Section,  
can be strongly modified if photons do mix with axion-like particles (ALPs).  
Pseudoscalar 
ALPs couple with photons
 through the following effective Lagrangian \cite{Raffelt:1987im}
\begin{equation}
{\cal L}_{a\gamma} =
-\frac{1}{4} g_{a\gamma} F_{\mu\nu}\tilde{F}^{\mu\nu}a
\label{eq:pscoupl}\,,
\end{equation}
where $a$ is the ALP field with mass $m_a$, $F_{\mu\nu}$ the
electromagnetic field-strength tensor, $\tilde{F}_{\mu\nu}
\equiv\frac{1}{2}\epsilon_{\mu\nu\rho\sigma}F^{\rho\sigma}$ its dual, 
and $g_{a\gamma} $ the ALP-photon coupling.
As a consequence of this coupling, ALPs and photons do oscillate into each other
in an external magnetic field. 
For a scalar particle, the coupling is proportional to
$F_{\mu\nu}F^{\mu\nu}a$. For definiteness, we limit our discussion to
the pseudoscalar case, but similar consequences apply also to scalars.

Let us suppose that a photon with energy $E$ moves in the $x_3$ direction.
 The transverse component of the external magnetic field is
 ${\bf B}_T={\bf B}-B_3{\bf e}_3$. The evolution equations of
 the photon-ALP system in presence of mixing and absorption 
are \cite{Raffelt:1987im,Csaki:2003ef} 
\begin{eqnarray}
i\frac{\partial}{\partial x_3}\left(\begin{array}{c}A_1\ \\ A_2 \\ a\end{array}\right)
&=&{\cal H}\left(\begin{array}{c}A_1\ \\ A_2 \\ a\end{array}\right) \nonumber\\
&=&
\left[\begin{array}{ccc}
\Delta_{11}-i\frac{\Gamma_\gamma}{2} & \Delta_{12} & \Delta_{a\gamma} c_\phi\\
\Delta_{21} & \Delta_{22}-i\frac{\Gamma_\gamma}{2} & \Delta_{a\gamma} s_\phi\\
\Delta_{a\gamma} c_\phi & \Delta_{a\gamma} s_\phi& \Delta_a\end{array}\right]
\left(\begin{array}{c}A_1\ \\ A_2 \\ a\end{array}\right)\, ,
\label{eq:evol1}\end{eqnarray}
where $c_\phi\equiv \cos\phi={\bf B}_T\cdot{\bf e}_1/B_T
= \sqrt{1-s_\phi^2}$.
 The entries $\Delta_{ij}$ ($i,j=1,2$) that
mix the photon polarization states are energy-dependent terms
determined by the properties of the medium and the QED vacuum
polarization effect. In particular,
neglecting the Faraday rotation effects
which are not relevant for the high energies of our interest,
 they read
\begin{eqnarray}
\Delta_{11}&=&\Delta_\parallel c^2_\phi+\Delta_\perp s^2_\phi\,,\nonumber\\
\Delta_{22}&=&\Delta_\parallel s^2_\phi+\Delta_\perp c^2_\phi\,,\nonumber\\
\Delta_{12}&=&\Delta_{21}=(\Delta_\parallel-\Delta_\perp)s_\phi c_\phi\,,
\end{eqnarray}
with $\Delta_\parallel=\Delta_{\rm pl}+\frac{7}{2}\Delta_{\rm QED}$, $\Delta_\perp=\Delta_{\rm pl}+2\Delta_{\rm QED}$ and
\begin{eqnarray}  
\Delta_{a\gamma}&=&\frac{1}{2} g_{a\gamma} B_T
\simeq   1.52\times10^{-2} \left(\frac{g_{a\gamma}}{10^{-11}\textrm{GeV}^{-1}} \right)
\left(\frac{B_T}{10^{-9}\,\rm G}\right) {\rm Mpc}^{-1}
\nonumber\,,\\  
\Delta_a&=&-\frac{m_a^2}{2E}
\simeq 
       -7.8 \times 10^{-4} \left(\frac{m_a}{10^{-10} 
        {\rm eV}}\right)^2 \left(\frac{E}{{\rm TeV}} \right)^{-1} {\rm Mpc}^{-1}
\nonumber\,,\\  
\Delta_{\rm pl}&=& -\frac{\omega^2_{\rm pl}}{2E}
 \simeq -1.1\times10^{-11}\left(\frac{E}{{\rm TeV}}\right)^{-1}
         \left(\frac{n_e}{10^{-7} \,{\rm cm}^{-3}}\right) {\rm Mpc}^{-1}
\nonumber\,,\\
\Delta_{\rm QED}&=& \frac{\alpha E}{45\pi}\left(\frac{B_T}{m_e^2/e}\right)^2
\simeq  
4.1\times10^{-9}\left(\frac{E}{{\rm TeV}}\right)
\left(\frac{B_T}{10^{-9}\,\rm G}\right)^2 {\rm Mpc}^{-1}
\,, 
\label{eq:Delta0}\end{eqnarray}
where 
$B_T$ is expressed in LorentzÐ-Heaviside units and
 $\omega^2_{\rm pl}=4\pi\alpha n_e/m_e$ is the plasma frequency of the medium,
 being $n_e$ the electron density. 
 For the numerical estimations above,
that we will use in the following as benchmark values,
 we have referred to the following physical input:
  The strength of
widespread, all-pervading $B$-fields in the extragalactic medium must
be 
$B \lesssim2.8\times10^{-7} (l/{\rm Mpc})^{-1/2}\,{\rm G}$,
coherent on a scale $l\simeq1\,$Mpc~\cite{Blasi:1999hu},
as obtained scaling the original bound from the Faraday effect of distant
radio sources~\cite{Kronberg:1993vk,Grasso:2000wj}
to the now much better known baryon density measured by the Wilkinson
Microwave Anisotropy Probe (WMAP)~\cite{Hinshaw:2008kr}.
The mean diffuse
intergalactic plasma density is bounded by $n_e \lesssim 2.7 \times
10^{-7}$~cm$^{-3}$, corresponding to the recent WMAP measurement of
the baryon density~\cite{Hinshaw:2008kr}.  Recent results from the
CAST experiment  give a direct experimental
bound on the ALP-photon coupling of $g_{\rm a\gamma}\lesssim 8.8\times 10^{-11}$~GeV$^{-1}$ for 
$m_a\lesssim 0.02$~eV~\cite{Arik:2008mq}, 
slightly better than the long-standing globular-cluster limit~\cite{Raffelt:2006cw}. 
 For ultra-light axions a stringent limit
from the absence of $\gamma$-rays from SN~1987A gives $\gag\lesssim
1\times 10^{-11}$~GeV$^{-1}$~\cite{Brockway:1996yr} or even $\gag
\lesssim 3\times 10^{-12}$~GeV$^{-1}$~\cite{Grifols:1996id}.
 Previous bounds on ALPs can be relaxed if they have a chameleontic
nature~\cite{Brax:2007ak}. In this case,  the best constraint
 comes from the structure of starlight polarization:
$g_{a \gamma} \lesssim 10^{-9}$~GeV$^{-1}$~\cite{Burrage:2008ii}.

The absorption term  $\Gamma_\gamma$ in Eq.~(\ref{eq:evol1}),
due to the VHE photons scattering with
the low-energy photons in the background,
produces a damping of the oscillations  in analogy 
with the case of the mixing of 
high-energy neutrinos in an absorbing matter~\cite{Naumov:2001ci}.
In the presence of this term, the Hamiltonian ${\cal H}$ 
is no longer hermitian. 

\subsection{Mean photon  transfer function}

Assuming an homogeneous magnetic field in a domain of size $l$,
in absence of absorption
the probability that a photon will convert into an 
ALP reads~\cite{Raffelt:1987im}
%%%%%%%%%%%%%%%%%%%%%%%%%%%%%%%%%%%%%%%%%%%%%%%%%%%%
\begin{equation}
\label{a16}
P_{a \gamma} = {\rm sin}^2 2 \theta \  {\rm sin}^2
\left( \frac{\Delta_{\rm osc} \, l}{2} \right)~,
\end{equation}
where the photon-ALP mixing angle $\theta$ is
\begin{equation}
\label{a16m}
\theta = \frac{1}{2} \, {\rm arcsin} \left( 
\frac{2 \Delta_{a \gamma}}{{\Delta}_{\rm osc}} \right) \,\ ,
\end{equation}
and the oscillation wavenumber reads
\begin{equation}
\label{a17}
{\Delta}_{\rm osc} = 
\left[\left( \Delta_a-\Delta_{\rm pl} \right)^2 + 
4 \Delta_{a \gamma}^2 \right]^{1/2}
= 2 \Delta_{a \gamma} \sqrt{1 + \left(\frac{E_c}{E} \right)^2}
 \,\ ,
\end{equation}
in terms of  the  critical energy
\begin{eqnarray}
{E}_c &\equiv& E
\frac{|\Delta_a-\Delta_{\rm pl}|}{2 \Delta_{a \gamma}}\nonumber\\
&\simeq&  2.5 \cdot 10^{-2} 
\frac{| m_a^2 - {\omega}_{\rm pl}^2|}{(10^{-10}{\rm eV})^2}
\left( \frac{10^{-9}{\rm G}}{B_T} \right)
\left( \frac{10^{-11}\rm GeV^{-1}}{g_{a \gamma}} \right)
{\rm TeV} \,\ .
\end{eqnarray}
In the high-energy  limit $E\gg E_c$,  $\Delta_{\rm osc}
\simeq 2 \Delta_{a \gamma}$, the  photon-ALP mixing is {maximal} ($\theta \simeq \pi/4$) 
and the conversion probability becomes energy-independent. 
This is the so-called strong-mixing regime. In this case,
if $\Delta_{a \gamma} l\ll 1$, the conversion probability 
on a single domain becomes extremely simple,
namely $P_{a \gamma} = (\Delta_{a \gamma}l)^2$.
In the following, we will work in this regime.

In this situation,  we explicitly drop  $\Delta_{\parallel,\perp}$ and $\Delta_a$ from 
the equations of motion [Eq.~(\ref{eq:evol1})].
 Thus, the propagation hamiltonian ${\cal H}$ can be written as
 ${\cal H}=\Delta-i{\cal D}$ where
\begin{equation}
\Delta = \Delta_{a\gamma}\left[\begin{array}{ccc}
0 & 0 & c_\phi\\
0 & 0 & s_\phi\\
c_\phi & s_\phi& 0\end{array}\right]\, ,
\label{eq:Delta}\end{equation}
and ${\cal D}=\frac{\Gamma_\gamma}{2}{\rm diag}(1,1,0)$ is the damping term
associated to the absorption.

Very high-energy gamma-rays propagate in the extragalactic magnetic fields
during their route to the Earth. These $B$-fields  
presumably have a turbulent structure. Therefore, 
for the case under study
 we need to describe photon-ALP conversions 
in random magnetic field configurations.
Let us now consider the propagation
of  photons  in many domains of  equal size $l$
($\simeq 1$~Mpc in our case)
  in which the magnetic field has
 (constant) random values and directions.  Along a given line of sight, the
 angles $\phi$ are randomly distributed in $[0,2\pi)$.
 In the following, we will work in the formalism 
of the density matrix
%...........................................
\begin{equation}
\rho = \left(\begin{array}{c}A_1\ \\ A_2 \\ a
\end{array}\right)
\otimes \left(\begin{array}{c}A_1 \  A_2 \ a\end{array}\right)^{*} \,\ .
\end{equation}
%%%%%%%%%%%%%%%%%%%%%%%%%%%%%%%%%%%%%%%%%%%%%%%%%%%% 
For the $k$-th domain the density  matrix is given by
\begin{equation}
{\rho}_k=e^{-i{\cal H}_kl}\cdot {\rho}_{k-1}\cdot e^{i{\cal H}_k^\dag l}\,,
\end{equation}
where $e^{-i{\cal H}_kl}$ is the propagation operator 
 for the $k$-th domain. During their path with a total length $L$, photons cross
$k=1,\ldots n$ domains ($n=L/l$) representing  a given random realization of  $B_k$ and $\phi_k$.
Since we cannot know this particular configuration, we  
perform an ensemble average over all the possible realizations on the
 $1,\ldots n$ domains. Defining this ensemble average as
 ${\bar \rho}_n=\langle{\rho}_n\rangle_{1\ldots n}$, we have
\begin{equation}
{\bar \rho}_n=\langle e^{-i{\cal H}_n l}\cdot \rho_{n-1}\cdot e^{i{\cal H}_n^\dag l}\rangle_{1\ldots n}=\langle e^{-i{\cal H}_n l}\cdot {\bar \rho}_{n-1}\cdot e^{i{\cal H}_n^\dag l}
\rangle_n\,.
\label{eq:average}\end{equation}
For the chosen values of the input parameters as in Eq.~(\ref{eq:Delta0}), we can
 perform a perturbative expansion up to the second order of the evolution operator in each domain, i.e.  

\begin{equation}
e^{-i{\cal H}_n l}\simeq 1-i{\cal H}_nl-\frac{1}{2} {\cal H}_n^2l^2
\,.\end{equation}

Performing then the ensemble average, as shown in Appendix A, 
using that  ${\bar \rho}_n-{\bar \rho}_{n-1}\simeq l\partial_{x_3} {\bar \rho}(x_3)$,
and summing over the two indistinguishable photon polarization states,
 we finally arrive at a system of two coupled differential equations
%%%%%%%%%%%%%%%%%%%%%%%%%%%%%%%%%%%%%%%%%%%%%%%%%%%%%%%%%%%%%%%%%%%%%%%%%%
\begin{equation}
\frac{\partial}{\partial x_3}
\left(\begin{array}{c} T_\gamma \\ T_a\end{array}\right)
=\frac{P_{a\gamma}}{l}\left[\begin{array}{cc}
-\left(\alpha+\frac{1}{2}\right) & 1\\
\frac{1}{2} & -1\end{array}\right]
\left(\begin{array}{c} T_\gamma\\ T_a\end{array}\right)\, ,
\label{eq:Tevol}\end{equation}
%%%%%%%%%%%%%%%%%%%%%%%%%%%%%%%%%%%%%%%%%%%%%%%%%%%%%%%%%%%
where $T_\gamma={\bar\rho}_{11} +{\bar\rho}_{22}$
and $T_a = {\bar\rho}_{aa}$ are the mean transfer functions for the
photon and for the ALP respectively;
$P_{a\gamma}={\Delta_{a\gamma}^2} l^2$ is
 the average photon-ALP conversion probability in each domain
 (in absence of absorption and in the limit of strong
mixing) and finally $\alpha= \Gamma_\gamma l/P_{a\gamma}$ is
the ratio between the absorption probability and the conversion probability. 

In realistic astrophysical situations
both $P_{a\gamma}$ and $\alpha$ are functions 
of the distance, due to the redshift dependence of
the extragalactic magnetic field and of the EBL.
However, taking these parameters as constant, 
in the hypothesis of only photons in the initial   state ($T_\gamma(0)=1$, $T_a(0)=0$)
Eq.~(\ref{eq:Tevol}) has a simple analytical solution 
\begin{equation}
T_\gamma(y) = e^{-\nu y}\left[\cosh \kappa y+\frac{1-2\alpha}{4\kappa}\sinh \kappa y\right]\, ,
\label{eq:Tgammaa}\end{equation}
where
\begin{eqnarray}
\nu &=&\frac{\alpha}{2}+\frac{3}{4}\, ,\nonumber\\
\kappa &=&\sqrt{\nu^2-\alpha}\, ,\nonumber\\
y &=& \frac{P_{a\gamma}x_3}{l} \,\ .
\end{eqnarray}
In particular, Eq.~(\ref{eq:Tgammaa}) gives the two limiting expressions
%.................................................................
\begin{equation}
\label{cases} T_\gamma(y) \simeq  \left\{
\begin{array}{ll}
 \frac{2}{3}+\frac{1}{3}e^{-3y/2} & \alpha =0, \\
\frac{1}{2\alpha^2}e^{-y}     & \alpha\gg 1. \\
\end{array}\right.
\end{equation}
%..................................................................... 
In the absence of absorption,  
we recover the mean  transfer function already  found in \cite{Grossman:2002by}. 
Conversely, in the case of strong absorption ($ \alpha\gg 1$)
 we have $T_\gamma\propto (\Gamma_\gamma)^{-2}$.
Using the approximate expression for $\Gamma_\gamma$ given in Eq.~(\ref{eq:bgktoday}) we observe that
the transfer function would drop as a power of the energy (rather than exponentially
as expected without ALP mixing).
Moreover, also the attenuation of the transfer function 
with the distance is less than in the case of absence of conversions. In fact, 
the argument of the exponential is suppressed by a factor $1/\alpha$ with respect
to the no-conversion case [see Eq.~(\ref{eq:noconversion})].

In Appendix B we report also the calculation of the root mean square
 $\delta T_\gamma$  for the distribution of 
the transfer function in different random realizations of the
magnetic field. This result is useful to estimate
the uncertainty associated with the averaging procedure. 

%%%%%%%%%%%%%%%%%%%%%%%%%%% FIGURE 2 %%%%%%%%%%%%%%%%%%%%%%%%%%%%%%%%%
\begin{figure}[!t]
\centering
\epsfig{figure=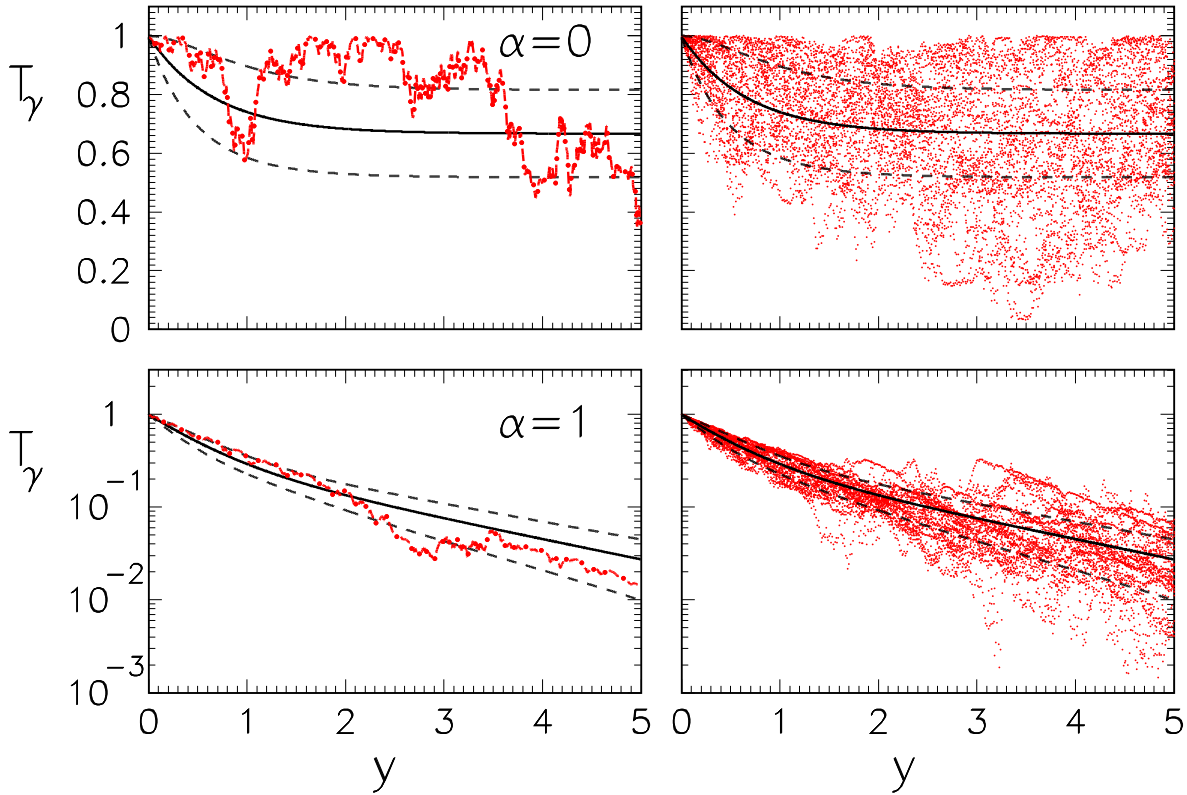, width =1.0\columnwidth, angle=0}
 \caption{Photon transfer function $T_{\gamma}(y)$ for 
$\alpha=0$ (no absorption) in the upper panels and for $\alpha=1$
in the lower panels. The continuous curve corresponds to the mean
value obtained by Eq.~(\ref{eq:Tgammaa}), while  the dashed lines
correspond to the dispersion around the mean $T_{\gamma}\pm\delta T_{\gamma}$, calculated
in Appendix~B. In the left panels, the dot-dashed lines correspond to a given 
realization
of the random magnetic fields along the photon line of sight,
obtained by the numerical integration of Eq.~(\ref{eq:evol1}). 
In the right panels we  show a scatter plot for  $T_{\gamma}$
   corresponding to $N_r=20$ realizations of the random magnetic
fields. (see the text for details)
\label{transfer}} 
\end{figure}
%%%%%%%%%%%%%%%%%%%%%%%%%%%%%%%%%%%%%%%%%%%%%%%%%%%%%%%%%%%%%%%%%%%%% 

For illustrative purposes, in Fig.~\ref{transfer} we compare the transfer matrix 
$T_{\gamma}(y)$ of Eq.~(\ref{eq:Tgammaa}),
with the numerical solution of Eq.~(\ref{eq:evol1}), for a value of the 
parameter $\Delta_{a\gamma}$  as in Eq.~(\ref{eq:Delta0}). For simplicity,
we choose constant values of the absorption factor
$\alpha$, namely $\alpha=0$ (no absorption) in the upper panel and 
$\alpha=1$ in the lower panel. The dashed lines represent 
$T_\gamma \pm \delta T_\gamma$, where $\delta T_\gamma$ 
is the dispersion calculated as in Appendix~B. 
In the left panels, the dot-dashed lines correspond to $T_{\gamma}$
for a given random realization
of the magnetic field  along the photon line of sight,
obtained by the numerical solution of Eq.~(\ref{eq:evol1}). In
 absence of absorption, 
we realize that along a given line of sight, the photon
transfer function can present strong deviations with respect to the
 average value obtained with our analytical calculation.  
 In this case, the  dispersion   with respect to the average tends to
$\delta T_\gamma=1/3\sqrt{5}$ for $y\to \infty$, as shown in Appendix~B. 
 We stress that the presence of a dispersion around the 
 average $T_{\gamma}$ has not been properly appreciated in the previous 
 literature. In particular, previous studies on the mixing  of photons with ALPs 
 emitted from point-like sources  into random magnetic fields
 have presented $2/3$ as limiting
value for the  photon transfer function (see e.g.~\cite{Csaki:2001yk} for the
case of photons emitted by supernovae Ia). 
In the case of strong mixing, where the oscillations are achromatic and
photons of different energies are not dephased, 
 this result is correct only on average, while
along a given line of sight one can expect ${\cal O} (1)$ deviations
from this value.
This peculiar effect has been also recently recognized in the context of photons emitted by
active galactic nuclei~\cite{Burrage:2009mj}. Indeed, the scatter in their luminosities has been interpreted
in~\cite{Burrage:2009mj} as a possible hint of the existence of a very light ALP. As a further consequence of this
effect,
 it would be worthwhile to investigate if  the presence of these large variations in the photon transfer function could 
put  additional constraints to the mechanism of photon-ALP conversions introduced in~\cite{Csaki:2001yk} to explain
the observed dimming of supernovae Ia. In fact,  photon conversions into ALP could produce large dispersions
in the supernova lightcurves.
 In the presence of
 absorption (lower panels)
the transfer function (and its dispersion) is suppressed as a power-law for $y\gg 1$, due 
to the damping of the oscillations. 
  In the right panels, we  superimpose on  our prescription for $T_{\gamma}$
 a scatter plot corresponding to $N_r=20$ different realizations of the random magnetic
fields 
and for 500 steps in the variable $y$. Again, we realize that behaviour of the $T_{\gamma}$ 
distribution is damped  by the effect of the absorption ($\alpha=1$).
 However, as we will see in the next Section,
 the presence of a dispersion around the mean value will have interesting consequences for VHE photons.
Finally, we mention that performing the averaging procedure over the transfer functions corresponding
 the different random realizations, we recover
our analytical results (not shown). 

\section{Transfer function for very high-energy photons}

We will  apply the  results of the previous Section to the case of VHE photons. 
 For the characterization of the  EBL,
 we  refer to   the 2008 Minimal Kneiske Model \cite{Kneiske:2008mp},
discussed in Section~2.
Concerning the extragalactic magnetic field, we consider it frozen into
the medium. With this assumption, the scaling
law of the magnetic field with the redshift $z$
is given by $B(z) = B_{0}(1+z)^2$~\cite{Grasso:2000wj}
while the size of the magnetic domains scales as  $l=l_0/(1+z)$.
With this choice, $P_{a\gamma}(z)$ scales as $(1+z)^2$. 
We assume $B_0=1$~nG and $l_0=1$~Mpc.
Equation~(\ref{eq:Tevol}) can be easily written in terms of the redshift,
by means of the Jacobian
%%%%%%%%%%%%%%%%%%%%%%%%%%%%%%%%%%%%%%%%%%%%%%%%%% 
\begin{equation}
\frac{dx}{dz}=\frac{c}{H_0}\frac{1}{(1+z)
\sqrt{\Omega_\Lambda + \Omega_m(1+z)^3}}
\, .
\end{equation}
%%%%%%%%%%%%%%%%%%%%%%%%%%%%%%%%%%%%%%%%%%%%%%%%%%%%%%%%%%%%%%%%%%%%%%%%%%%%%

\begin{figure}[!t]
\centering
\epsfig{figure=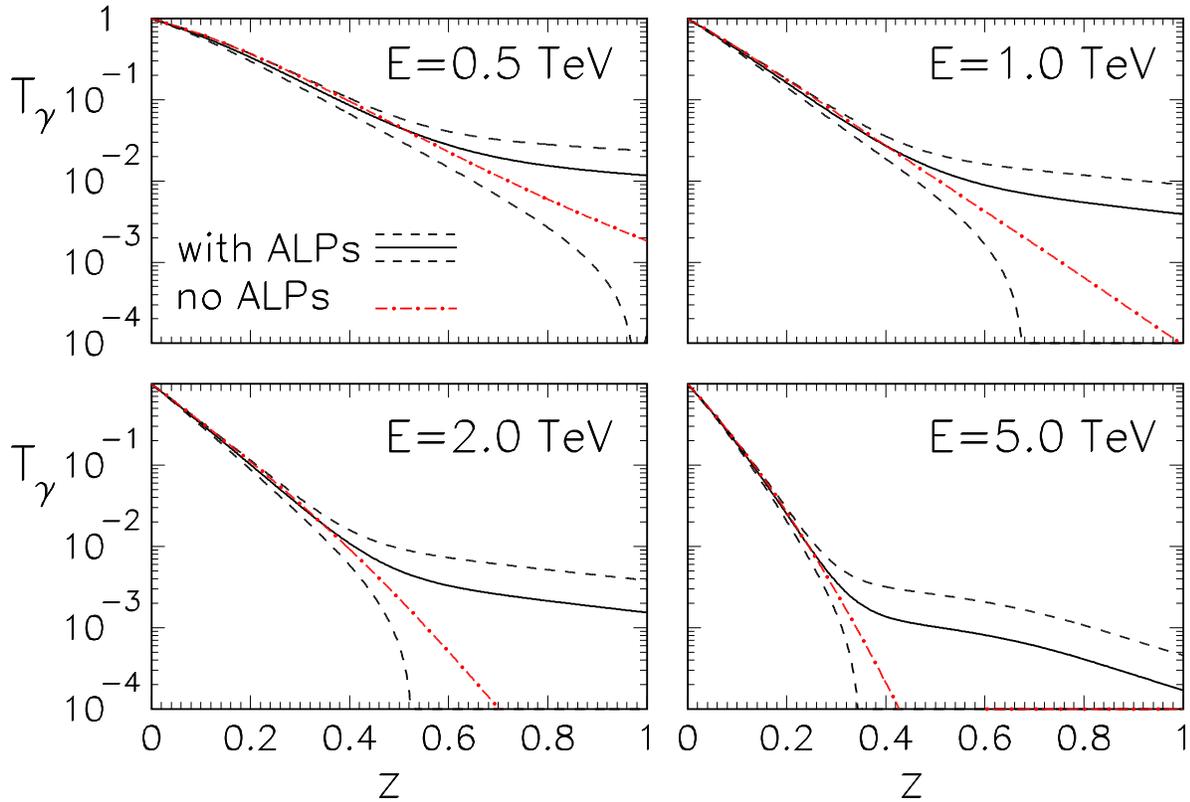, width =1.0\columnwidth, angle=0}
 \caption{ 
VHE photons transfer function $T_{\gamma}$ in function of the redshift $z$,
for different values of the observed photon  energy $E$. The standard case, including
only  photon absorption on the EBL, is represented by the dot-dashed curve.
In the presence  of   mixing with ALPs 
the mean photon transfer function, obtained by Eq.~(\ref{eq:Tevol}),
is represented by the solid curve and its typical spread $ T_{\gamma}
\pm \delta T_\gamma$
corresponds to the two dashed lines. 
\label{redshift}} 
\end{figure}

In Fig.~\ref{redshift} we show the photon transfer function $T_{\gamma}$
in presence of absorption on the EBL for our reference model, in function
of the redshift $z$ for different values of the  observed photon  energy  
$E$ with (continuous curve) and without ALP  mixing   (dotted curve).
The dashed curves represent the spread $T_{\gamma}
\pm \delta T_{\gamma}$ around the mean value.
One realizes that at redshift $z>0.2$ the presence of ALPs could  produce 
dramatic modifications in the shape of the photon transfer function,
the stronger the effect the higher the photon energy. 
In this sense, a hint   of the mixing of very  high-energy photons with  ALPs
 would be the detection of very distant sources otherwise obscure due to  the absorption.
However, from the spread of the  photon transfer function $\delta T_{\gamma}$, we
also realize
that at high redshift $T_{\gamma}$ can present relevant deviations with  
 respect to the mean value. In this sense, the effect of  mixing with ALPs for 
VHE photons emitted by  distant
gamma sources would be strongly dependent on the particular realization of the
extragalactic magnetic fields crossed by them during their propagation. 
Therefore, it  is not  guaranteed that the mixing  with ALPs
could provide an universal mechanism to achieve the transparency
of distant gamma-sources. Conversely, for particular configurations of
the extragalactic magnetic fields crossed by VHE photons, one could
also find  a  suppression of the transfer function 
stronger than in the standard case. 
In general,  observing VHE photons  from far sources 
one would  find strong differences in the measured spectra
along different lines of sight.   
As a further consequence, the presence of ALPs would prevent the possibility
to use distant gamma sources at very high energy as cosmological candles~\cite{Bi:2008rf}.

%%%%%%%%%%%%%%%%%%%%%%%%%%% FIGURE 3 %%%%%%%%%%%%%%%%%%%%%%%%%%%%%%%%%
%%%%%%%%%%%%%%%%%%%%%%%%%%%%%%%%%%%%%%%%%%%%%%%%%%%%%%%%%%%%%%%%%%%%% 

\begin{figure}[!t]
\centering
\epsfig{figure=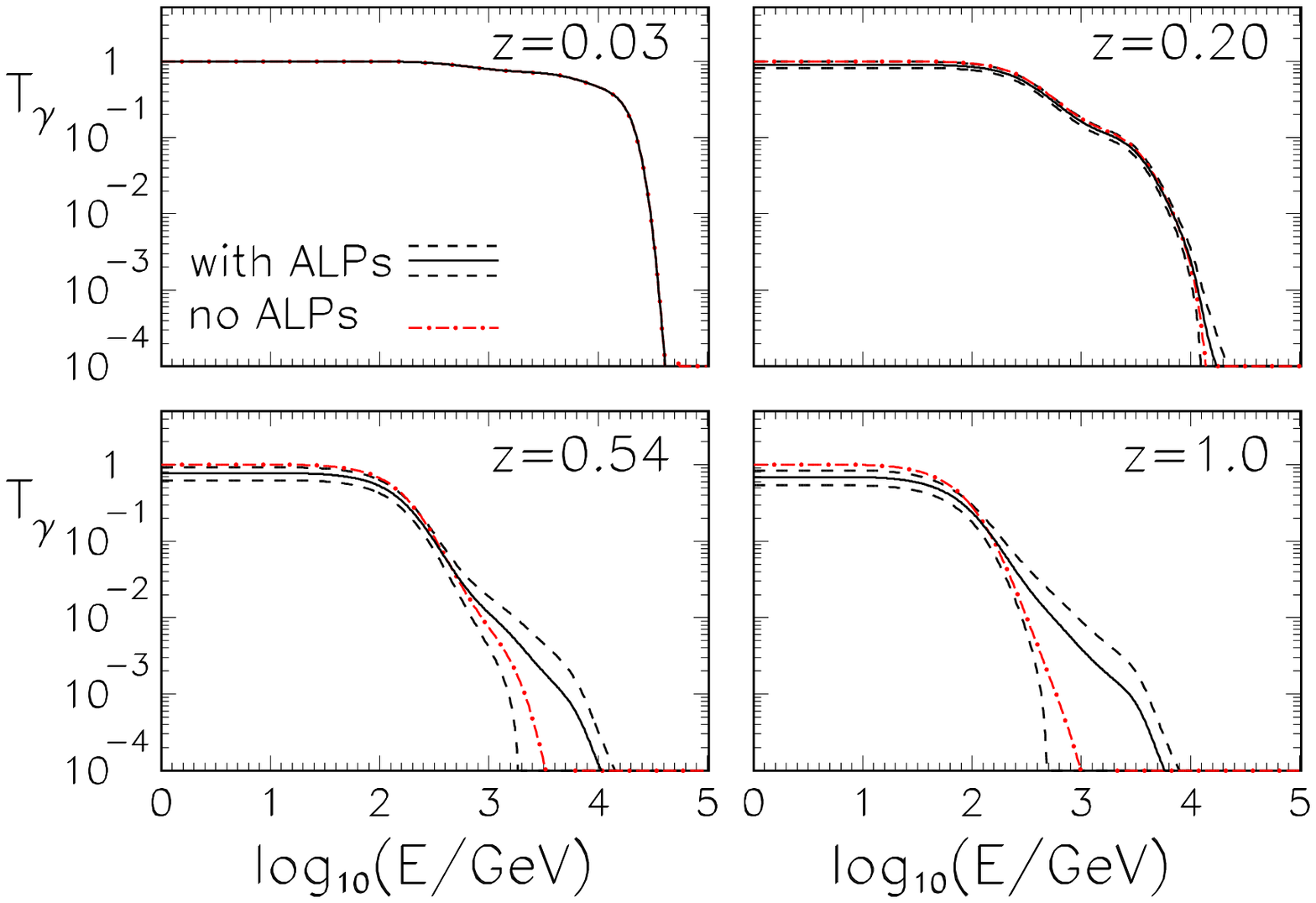, width =1.0\columnwidth, angle=0}
 \caption{ VHE photons transfer function $T_{\gamma}$ 
in function of the observed photon energy $E$,
for different values of the redshift $z$.
The labels of the different curves are the same as in Fig.~\ref{redshift}.
\label{energy}} 
\end{figure}

In Fig.~\ref{energy} we show the photon transfer function as function
of the energy for four different values 
of the redshift of the emitting 
source (as in Fig.~1 of~\cite{Kneiske:2008mp}). We see that in absence of ALP conversions the photon transfer function 
would be strongly suppressed at energies above $E \gtrsim 100$~GeV, the stronger the 
suppression the larger the redshift. 
On the other hand, in presence of conversions, $T_\gamma$ has an approximate power-law behaviour at high energies, as pointed out in the previous Section.
We also realize that the spread in the possible values  of $T_{\gamma}$
would make difficult to infer strong conclusions about ALP mixing
observing only few sources. 
We also note that in the case of $z=0.20$ the inclusion of the ALPs does not produce any significant
change in the photon transfer function. This  suggests that it would be difficult to interpret
in terms of ALP conversions the presumed
transparency to gamma radiations for the sources at $z=0.165$ and $z=0.186$ discussed in~\cite{Aharonian:2005gh,Mazin:2007pn}.
Conversely, ALP conversions could play a significant role for the source 3C279 at redshift $z=0.54$~\cite{alberto}.

Finally, in Figure~5  we show two iso-contours of the photon transfer function $T_{\gamma}$ in function
of the redshift $z$ and of the observed photon energy $E$ with  and
without ALP conversions. 
We see that the attenuation of the photon transfer function
in the presence of ALP conversions  could present a large variation
 at high redshift with respect to the standard expectations with only absorption.

%%%%%%%%%%%%%%%%%%%%%%%%%%% FIGURE 4 %%%%%%%%%%%%%%%%%%%%%%%%%%%%%%%%%
%%%%%%%%%%%%%%%%%%%%%%%%%%%%%%%%%%%%%%%%%%%%%%%%%%%%%%%%%%%%%%%%%%%%% 

%%%%%%%%%%%%%%%%%%%%%%%%%%% FIGURE 5 %%%%%%%%%%%%%%%%%%%%%%%%%%%%%%%%%
\begin{figure}[!t]
\centering
\epsfig{figure=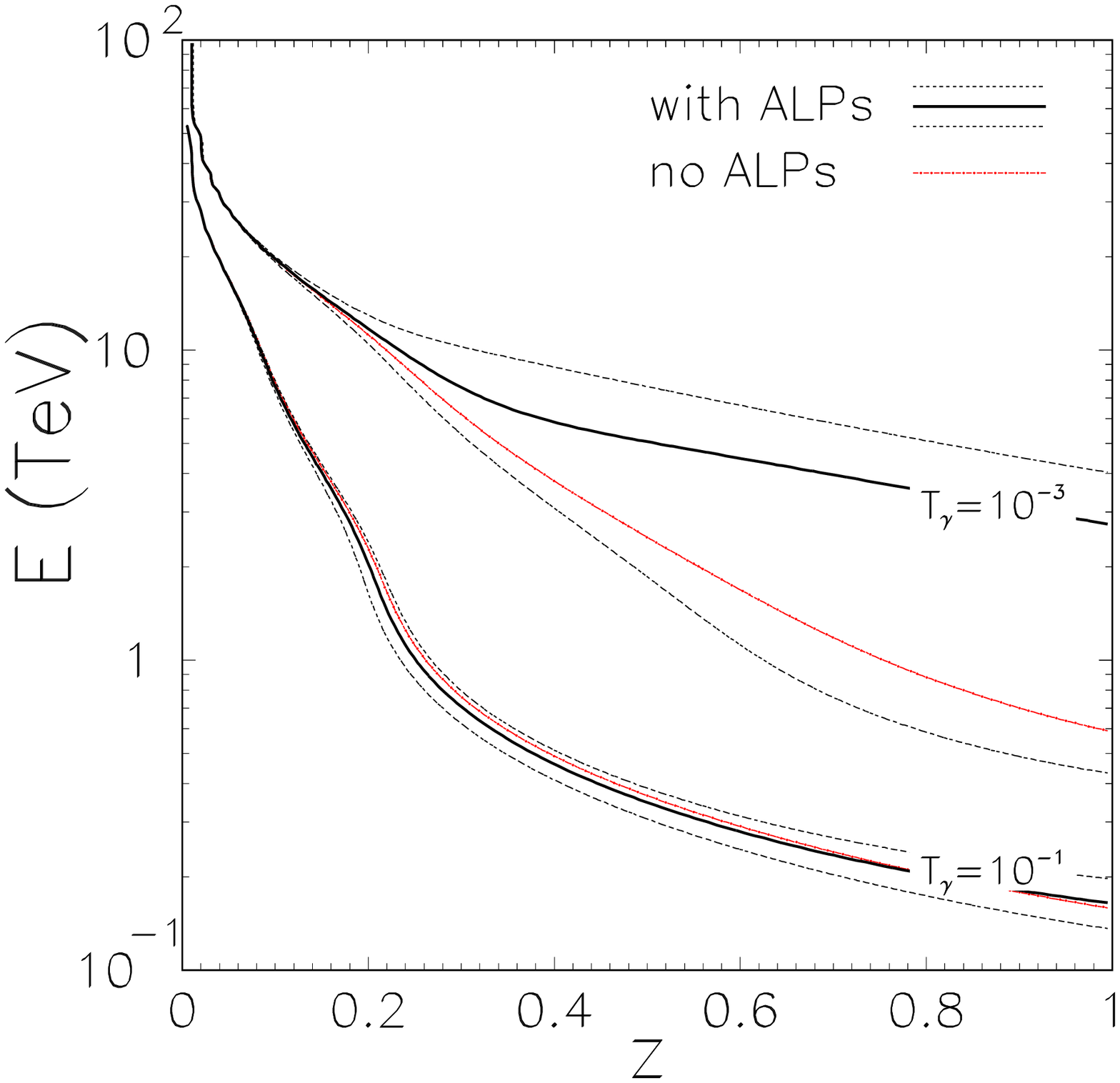, width =1\columnwidth, angle=0}
 \caption{Iso-contours of the photon transfer function $T_{\gamma}$ in function
of the redshift $z$ and of the observed photon energy with only absorption (light  curves) and
in presence of ALP conversions. In this latter case, we represent the mean value of $T_{\gamma}$ (black solid curves)
and its dispersion $T_{\gamma}\pm\delta T_{\gamma}$ (black dashed curves).}
\label{cont}
\end{figure}
%%%%%%%%%%%%%%%%%%%%%%%%%%%%%%%%%%%%%%%%%%%%%%%%%%%%%%%%%%%%%%%%%%%%% 

\section{Discussion and conclusions}

Very high-energy gamma-ray observations would open the possibility to probe
the existence of  axion-like particles (ALPs) predicted in many theories
beyond the Standard Model. 
Recent gamma observations 
of cosmologically distant gamma-ray 
sources  have
 revealed a surprising degree of transparency of the universe
 to very high-energy photons.
The oscillations between  high-energy  photons and ALPs
in the random extragalactic magnetic fields 
have been proposed as an 
intriguing possibility to explain these 
observations~\cite{De Angelis:2007dy,DeAngelis:2008sk}. 
Apart from the original proposal,
the consequences of this mechanism are testable with the measurements
of the new generation of Imaging Atmospheric  Cherenkov Telescopes, like
MAGIC~\cite{magic}, HESS~\cite{hess},
 VERITAS~\cite{Weekes:2001pd} or CANGAROO-III~\cite{Enomoto:2001xu},
 covering energies in the range
0.1-20~TeV, and hopefully with the future Cherenkov Telescope Array, reaching
energies of 100~TeV~\cite{cta}.

In order to perform a systematic study of these effects, without recurring
to brute-force time-consuming numerical simulations,
in this paper 
 we have presented
a simple calculations of the  photon mean transfer function 
and of its variance in presence of
absorption on the EBL and mixing with ALPs. 
We have found that our prescription is enough accurate for the case under
study and we have shown some numerical examples of VHE photon transfer functions.
It results that  VHE photon-ALP mixing would produce peculiar
deformations in the energy spectra of very high-energy gamma-rays emitted at 
 high redshift. We have also found that the photon transfer function
 in presence of  VHE photon-ALP conversions
 presents a relevant dispersion around the mean value  due
to the randomness of the extragalactic magnetic fields crossed by
the photons.  Due to this fact, the measured  flux 
for VHE photons traveling along different lines of sight can be 
strongly suppressed or enhanced  with respect
to the case with only absorption, depending on the particular
configuration of random magnetic fields encountered.
This would suggest that photon-ALP conversions could not be an universal
mechanism to produce the transparency of universe to VHE photons.
Conversely, 
the most striking signature of the mixing  with
ALPs would be a reconstructed EBL density
from TeV photon observations which appears to vary over different directions
of the sky: consistent
with standard expectations in some regions, but inconsistent in others.
To test this effect  we would  need to collect data from  sources along different
directions in the sky in order to perform a study of the
photon energy distributions, from which we could hope to  infer possible hints of ALPs. 
A further signature of these stochastic conversions would be the detection
of peculiar direction-dependent dimming effects in the diffuse photon radiation observable in GeV range,
testable with the FERMI (previously called GLAST) experiment~\cite{Gehrels:1999ri}.

As further developments, we plan to use our calculation to  perform a systematic study of ALP
signatures in very high-energy gamma-rays, analyzing in details the spectral deformations
 expected for 
observed sources at different redshifts
 and for different models of the extragalactic background light. 
 Thanks to our calculation,
this task now appears more doable than before. 
After that, it will remain  to see if  elusive  ALPs will show up from the sky.

\section*{Acknowledgements}

We thank R.\ Wagner and D.\ Shaw for fruitful discussions and
 C.\ Burrage, G.\ Raffelt, M.\ Roncadelli and P.\ D.\ Serpico for reading
the manuscripts and for useful comments on it. 
D.M.\ acknowledges
kind hospitality at the Max-Planck-Institut where part of this work was done. D.M. 
thanks also the organizers of the ``5th Patras Workshop on Axions, WIMPs and WISPs''
for the kind hospitality and the stimulating discussions.
In Lecce, the work of D.M.\ is partly supported  by the Italian MIUR and INFN through the 
``Astroparticle Physics'' research project, and by the EU ILIAS through the 
ENTApP project.

\appendix

\section{Calculation of the mean photon transfer function}

Here, we  present the derivation of the photon transfer function $T_{\gamma}$
introduced in Sec.~3.2.
We have defined  
 ${\bar \rho}_n\equiv\langle{\rho}_n\rangle_{1\ldots n}$, 
as an ensemble average of the density matrix $\rho_n$  
 in the $n$-th domain 
over all the possible realizations of the random magnetic fields in all the domains from $1$ to $n$. Since ${\cal H}_n$ depends only from the configuration of the $n$-th domain, we have that 
\begin{equation}
{\bar \rho}_n=\langle e^{-i{\cal H}_n l}\cdot \rho_{n-1}\cdot e^{i{\cal H}_n^\dag l}\rangle_{1\ldots n}=\langle e^{-i{\cal H}_n l}\cdot {\bar \rho}_{n-1}\cdot e^{i{\cal H}_n^\dag l}
\rangle_n\, .
\label{eq:averageap}\end{equation}
For the chosen values of the input parameters  in Eq.~(\ref{eq:evol1}), we can perform a perturbative expansion of the evolution operator up to the 
 second order
${\cal H}_n = \Delta_n -i {\cal D}_n$ in each domain, i.e.  
\begin{equation}
e^{-i{\cal H}_n l}\simeq 1-i{\cal H}_n l-\frac{1}{2} {\cal H}^2_n l^2
\,.\end{equation}
Since, from Eq.~(\ref{eq:Delta}) it results that
$\langle \Delta_n\rangle_{\phi_n}=0$, 
 Eq.~(\ref{eq:averageap}) can be written as
\begin{eqnarray}
{\bar\rho}_n &=&{\bar \rho}_{n-1}-l\left({\cal D}_n{\bar \rho}_{n-1}+
{\bar \rho}_{n-1}{\cal D}_n\right) \nonumber \\
&+&
l^2\langle \Delta_n {\bar \rho}_{n-1} \Delta_n \rangle_n
 -\frac{l^2}{2}\left(\langle \Delta_n^2\rangle_n {\bar \rho}_{n-1}
+{\bar \rho}_{n-1}\langle \Delta_n^2\rangle_n\right)\, .\label{eq:eqQ}
\end{eqnarray}
In the previous equation we have neglected the second order terms in 
${\cal D}_n l$ since they give a contribution of the order of 
$(\Gamma_\gamma l)^2\lesssim 10^{-6}$ [see Eq.~(\ref{eq:lenght})] which is 
at least two orders of magnitude smaller than  $(\Delta_{a\gamma} l)^2\sim 10^{-4}$, 
when we use the benchmark values in Eq.~(\ref{eq:Delta0}).

For $\langle \Delta_n^2\rangle_n$ we have
\begin{eqnarray}
\langle \Delta_n^2\rangle_n&=&\frac{1}{4} g^2_{a\gamma}\left\langle B_{T,n}^2
\left[\begin{array}{ccc}
\cos^2\phi_n & \sin\phi_n\cos\phi_n & 0 \\
\sin\phi_n\cos\phi_n & \sin^2\phi_n & 0 \\
0 & 0 & 1
\end{array}\right]\right\rangle_{\phi_n, {\bf B}_n}\nonumber\\
&\equiv& 
\overline{\Delta_{a\gamma}^2}
\left[\begin{array}{ccc}
\frac{1}{2} & 0 & 0 \\
0 & \frac{1}{2} & 0 \\
0 & 0 & 1
\end{array}\right]\, ,
\end{eqnarray}
where we have defined 
\begin{equation}
\overline{\Delta^2_{a\gamma}}=\frac{1}{4}g^2_{a\gamma}\langle B_T^2\rangle_{{\bf B}_n}
=\left(\frac{1}{2}g_{a\gamma} B_{\rm eff}\right)^2\, ,
\end{equation}
where $ B_{\rm eff}^2 = \langle B_T^2\rangle_{{\bf B}_n} =2\langle|{\bf B}|^2\rangle/3$ due to the projection effect.
In the same way we have
\begin{equation}
\langle \Delta_n {\bar \rho}_{n-1}\Delta_n \rangle_n = 
\frac{1}{2}\overline{\Delta_{a\gamma}^2}
\left[\begin{array}{ccc}
{\bar\rho}_{ aa} & 0      & {\bar\rho}_{ a1}\\
0      & {\bar\rho}_{ aa} & {\bar\rho}_{ a2}\\
{\bar\rho}_{1a} & {\bar\rho}_{ 2a} & 
{\bar\rho}_{{11}}
+{\bar\rho}_{ 22}
\end{array}\right]_{n-1}\, .
\end{equation}
After  $n$ domains,
 the total distance travelled by the photon is $x_3=n l$. 
 Defining ${\bar \rho}_n\equiv {\bar \rho}(x_3)$,
 one has  ${\bar \rho}_n-{\bar \rho}_{n-1}\simeq l\partial_{x_3} {\bar \rho}(x_3)$.
After a straightforward 
calculation, we obtain  the evolution equation
 for the averaged density matrix ${\bar \rho}$
\begin{equation}
\nonumber
\frac{\partial {\bar \rho}}{\partial {x_3} }
=\frac{P_{a\gamma}}{l}
\left[\begin{array}{ccc}
-\mu{\bar \rho}_{11} + \frac{1}{2}{\bar \rho}_{aa} &
-\mu{\bar \rho}_{12} &
-\nu{\bar \rho}_{1a} + \frac{1}{2}{\bar \rho}_{a1}\\
-\mu{\bar \rho}_{21} &
-\mu{\bar \rho}_{22} + \frac{1}{2}{\bar \rho}_{aa} &
-\nu{\bar \rho}_{2a} + \frac{1}{2}{\bar \rho}_{a2}\\
-\nu{\bar \rho}_{a1} + \frac{1}{2}{\bar \rho}_{1a} &
-\nu{\bar \rho}_{a2} + \frac{1}{2}{\bar \rho}_{2a} &
   - {\bar \rho}_{aa} + \frac{1}{2}\left({\bar \rho}_{11}+{\bar \rho}_{22}\right)
\end{array}\right]\, 
\label{eq:Qevolapp}
\end{equation}
where 
\begin{eqnarray}
\mu &=& \alpha+\frac{1}{2} \,\ , \nonumber \\
\nu &=& \frac{\alpha}{2}+\frac{3}{4} \,\ ,
\end{eqnarray}
and we have  
 defined as
 $P_{a\gamma}=\overline{\Delta_{a\gamma}^2} l^2$ 
 the average photon-ALP oscillation probability in each domain
 (in absence of absorption and in the limit of strong
mixing) and  $\alpha\equiv \Gamma^\gamma l/P_{a\gamma}$ 
the ratio between the absorption probability and the oscillation probability. 
 
Since we are interested in determining the total final photon and ALP flux,
 we define the mean {\em transfer functions} 
$T_\gamma= {\bar\rho}_{11}+{\bar\rho}_{22}$ 
and $T_a={\bar\rho}_{aa}$, for whose evolution we finally obtain
\begin{equation}
\frac{\partial}{\partial x_3}
\left(\begin{array}{c} T_\gamma \\ T_a\end{array}\right)
=\frac{P_{a\gamma}}{l}\left[\begin{array}{cc}
- \mu & 1\\
\frac{1}{2} & -1\end{array}\right]
\left(\begin{array}{c} T_\gamma\\ T_a\end{array}\right)\, .
\label{eq:Tevolap}\end{equation}
Defining the variable $dy=P_{a\gamma} dx_3/l$, we can also obtain a second order equation for the function $T_\gamma$
\begin{equation}
T''_\gamma+\left(\alpha+\frac{3}{2}\right)T'_\gamma+\left(\alpha+\alpha'\right)T_\gamma=0\, ,
\label{eq:Teq}\end{equation}
where the prime denotes the derivative with respect to $y$. 
For constant $\alpha$ we easily obtain the solution in Eq.~(\ref{eq:Tgammaa}).

\section{Calculation of  higher  momenta }

In the previous Appendix
 we have have calculated
the photon transfer function $T_{\gamma}$ averaged 
 over all the possible magnetic domain configurations.
 However, photons coming from a single source cross just one (unknown) 
particular realization of the magnetic field domains.
 It is   thus interesting  to evaluate the uncertainty
 introduced by the procedure of averaging. To do this, we  calculate the second 
order momenta of the probability distributions, by means of a procedure similar to 
the one  introduced in Appendix~A. We define the ``square'' of the density matrix  
as
\begin{equation}
{\rho}^{(2)}={\rho}\otimes {\rho}\rightarrow \rho^{(2)}_{ijkl}=\rho_{ij}\rho_{kl}\, ,
\end{equation}
and ${\bar \rho}^{(2)}_n=
\langle {\rho}^{(2)}_n\rangle_{1\dots n}$.
 We rewrite Eq.~(\ref{eq:Qevolapp}) in the following way
\begin{equation}
\frac{\partial}{\partial y} {\bar \rho}_{ij}=G_{ijrs}{\bar\rho}_{rs}\, ,
\end{equation}
where the tensor $G_{ijkl}$ can be written as
\begin{displaymath}
G_{ijkl}=\left\{
\begin{array}{cllll}
-\mu && {\rm if} & ijkl= & 1111,\,1212,\,2121,\,2222
\\
-1 && {\rm if} & ijkl= & aaaa\\
-\nu && {\rm if} & ijkl= & 1a1a,\,a1a1,\,2a2a,\,a2a2\\
\frac{1}{2} &&{\rm if} & ijkl= & 11aa,\,22aa,\,aa11,\,aa22\\
& & & &1aa1,\,2aa2,\,a11a,\,a22a\\
0 & & & &{\rm otherwise}\, .
\end{array}
\right.
\end{displaymath}
Performing the  average of $\rho^{(2)}$
  up to the first order in $\Gamma_\gamma l$ and to the  second
 order in $\Delta_{a\gamma}l$ as in Eq.~(\ref{eq:eqQ}), we arrive
at the following equation for the evolution of ${\bar \rho}^{(2)}$
\begin{eqnarray}
\frac{\partial }{\partial x_3} {\bar\rho}^{(2)}_{ijkl}&=&
\frac{P_{a\gamma}}{l}\left(G_{ijrs}{\bar\rho}^{(2)}_{rskl}+G_{rskl}{\bar\rho}^{(2)}_{ijrs}\right)\nonumber\\
&\phantom{=}&-l\langle
\left(\Delta_{ir}\rho_{rj}-\rho_{ir}\Delta_{rj}\right)
\left(\Delta_{ks}\rho_{sl}-\rho_{ks}\Delta_{sl}\right)
\rangle_{1\cdots n}\nonumber\\
&=&
\frac{P_{a\gamma}}{l}\left(G_{ijrs}{\bar\rho}^{(2)}_{rskl}+G_{rskl}{\bar\rho}^{(2)}_{ijrs}\right)\nonumber\\
&\phantom{=}&-l\left(
\langle \Delta_{ir}\Delta_{ks}\rangle {\bar\rho}^{(2)}_{rjsl}+
\langle \Delta_{rj}\Delta_{sl}\rangle {\bar\rho}^{(2)}_{irks}\right.\nonumber\\
&\phantom{=}&-\left.
\langle \Delta_{ir}\Delta_{sl}\rangle {\bar\rho}^{(2)}_{rjks}-
\langle \Delta_{rj}\Delta_{ks}\rangle {\bar\rho}^{(2)}_{irsl}
\right)
\, ,
\label{eq:complete}\end{eqnarray}
where for simplicity  we have dropped the subscript $n$ from the averages
in the last two lines. The last term arises from the linear 
term in $l$ in Eq.~(\ref{eq:average}) which is absent in Eq.~(\ref{eq:eqQ})
 since it averages out ($\langle{\Delta}\rangle=0$). From Eq.~(\ref{eq:Delta}) we have that $\langle \Delta_{ij}\Delta_{kl}\rangle=\frac{1}{2}\overline{\Delta_{a\gamma}^2}\xi_{ijkl}$ with
\begin{displaymath}
\xi_{ijkl}=\left\{
\begin{array}{cllll}
1 && {\rm if} & ijkl= &11aa,\,1aa1,\,a11a,\,aa11\\
& & &                 &22aa,\,2aa2,\,a22a,\,aa22\\
0 & & & &{\rm otherwise}\, .
\end{array}
\right.
\end{displaymath}
After a long but straightforward derivation, 
   one can extract a subset of 6 independent equations
out from the set of the 81 of (\ref{eq:complete}):
\begin{eqnarray}
\partial_y R_\gamma &=& -(2\alpha+1)R_\gamma+2\eta_{a\gamma}-\zeta_{a\gamma}
\nonumber\\
\partial_y R_a &=& -2R_a + \eta_{a\gamma}-\zeta_{a\gamma}
\nonumber\\
\partial_y R_p &=& -(2\alpha+1) R_p -\zeta_{a\gamma}
\nonumber\\
\partial_y \zeta_\gamma &=& -(2\alpha+1) \zeta_\gamma -\zeta_{a\gamma}
\nonumber\\
\partial_y \eta_{a\gamma} &=& -\left(\alpha+\frac{3}{2}\right)\eta_{a\gamma}
+\frac{1}{2}R_\gamma+R_a+\zeta_{a\gamma}
\nonumber\\
\partial_y \zeta_{a\gamma} &=& -\left(\alpha+\frac{5}{2}\right)\zeta_{a\gamma}
-\frac{1}{2}R_\gamma-2R_a+2\eta_{a\gamma}-\frac{1}{2}(R_p+\zeta_\gamma)\, ,
\label{eq:6equations}\end{eqnarray}
where, as usual, $dy=P_{a\gamma}dx_3/l$ and,
\begin{eqnarray}
R_\gamma &=& {\bar\rho}^{(2)}_{1111}+{\bar\rho}^{(2)}_{2222}+{\bar\rho}^{(2)}_{1122}
+{\bar\rho}^{(2)}_{2211}
\equiv \langle ({\bar\rho}_{11}+{\bar\rho}_{22})^2\rangle
\nonumber\\
R_a &=& {\bar\rho}^{(2)}_{aaaa}\equiv \langle {\bar\rho}_{aa}^2\rangle
\nonumber\\
R_p &=& {\bar\rho}^{(2)}_{1111}+{\bar\rho}^{(2)}_{2222}-{\bar\rho}^{(2)}_{1122}-{\bar\rho}^{(2)}_{2211}
\equiv \langle ({\bar\rho}_{11}-{\bar\rho}_{22})^2\rangle
\nonumber\\
\zeta_\gamma &=& {\bar\rho}^{(2)}_{1212}+{\bar\rho}^{(2)}_{2121}+{\bar\rho}^{(2)}_{1221}+
{\bar\rho}^{(2)}_{2112}
\nonumber\\
\eta_{a\gamma} &=& \frac{1}{2}\left({\bar\rho}^{(2)}_{11aa}+{\bar\rho}^{(2)}_{aa11}
+{\bar\rho}^{(2)}_{22aa}+{\bar\rho}^{(2)}_{aa22}\right)\equiv\langle
 ({\bar\rho}_{11}+{\bar\rho}_{22})\cdot {\bar\rho}_{aa}\rangle
\nonumber\\
\zeta_{a\gamma} &=& \frac{1}{2}\bigg({\bar\rho}^{(2)}_{1a1a}+{\bar\rho}^{(2)}_{a1a1}-
{\bar\rho}^{(2)}_{a11a}-{\bar\rho}^{(2)}_{1aa1} \nonumber\\
&+&{\bar\rho}^{(2)}_{2a2a}
+{\bar\rho}^{(2)}_{a2a2}-{\bar\rho}^{(2)}_{a22a}-{\bar\rho}^{(2)}_{2aa2}\bigg)
\, .
\end{eqnarray}
We can thus give  a physical interpretation for some of these quantities: 
 $R_\gamma$ and $R_a$  are the square average of the photon and ALP transfer function
respectively, $R_p$ is the average square degree of polarization of the photon,
 $\eta_{a\gamma}$ is the photon-ALP correlation.
 Since the third and the fourth of Eqs.~(\ref{eq:6equations})
 are similar,  starting from a  completely unpolarized  photon 
state we find also that $R_p(y)=\zeta_\gamma(y)$, 
so that we can reduce further 
the system Eqs.~(\ref{eq:6equations}) to five independent equations.
%\footnote{Actually, with a further manipulation we can show that those system can be decomposed in two independent subsystems of 3+2 equations.}

We finally define the ``1$\sigma$'' uncertainty on the photon transfer function as
 $\delta T_\gamma=[R_\gamma-T_\gamma^2]^{1/2}$. However,  the 
distribution of $T_\gamma$ is not gaussian (and in general is also asymmetrical) 
so that $T_\gamma\pm\delta T_\gamma$ should be interpreted
 just as a qualitative ``band'' with a not defined confidence level. 

For a constant value of $\alpha$,  the system of Eqs.~(\ref{eq:6equations}) can
 be integrated analytically. For simplicity, we give here only the solution for the case $\alpha=0$
\begin{equation}
R_\gamma(y)=\frac{49+50e^{-3y/2}+6e^{-5y}}{105}\, .
\end{equation}
For $y\to\infty$ we have thus the prediction $T_\gamma=2/3\pm 1/3\sqrt{5}$.

\end{document}